\documentclass[prl,letterpaper,twocolumn,showpacs,superscriptaddress]{revtex4}
\usepackage{graphicx,psfrag,amsmath,amssymb,amsfonts,bbm,latexsym,color,dcolumn}

\begin{document}

\title{Electron-barrier interaction in a vacuum tunneling probe}

\author{Roberto Onofrio}

\affiliation{Dipartimento di Fisica ``G. Galilei'', Universit\`a di
  Padova, Via Marzolo 8, 35131 Padua, Italy}

\author{Carlo Presilla}

\affiliation{Dipartimento di Fisica, Universit\`a di Roma ``La
  Sapienza'', Piazzale A. Moro 2, 00185 Rome, Italy}

\date{8 June 1992, published in {\sl Phys. Lett. A} {\bf 166} (1992) 24}

\begin{abstract}
A model for dealing with energy and momentum exchanges between
ballistic electrons and the vacuum barrier in a tunneling probe used
as an electromechanical transducer is studied and its physical
significance in devices of size comparable to the mean free path of
the tunneling electrons is discussed.
\end{abstract}

\maketitle

The use of tunneling probes for scanning microscopy on surfaces is
well known (see for instance ref. \cite{PLA1}). More recently, use of
tunneling probes to monitor displacements of macroscopic masses has
been proposed as a high-sensitivity, low-noise electromechanical
transducer to detect gravitational waves \cite{PLA2}. Further
investigations on the device have shown that the back-action of the
amplifier following the transducer is negligible and that the quantum
limit comes earlier from the interaction process between the tunneling
electrons and the barrier \cite{PLA3}. The application of the
Heisenberg uncertainty principle to a vacuum tunneling probe has been
the subject of two papers in which calculations in a second
quantization \cite{PLA4} and in a first quantization framework
\cite{PLA5} have been performed. The underlying physical hypothesis is
the complete release of the momentum and energy of the tunneling
electrons to the test mass. However, in the same papers \cite{PLA4,PLA5}
a tunneling transducer is proposed to measure interatomic forces and
to detect quantum noise. Due to the small size of the test mass in
both these situations the total absorption of the tunneling electrons
is not assured. When the sample has a size smaller than the mean free
path for inelastic scattering the energy of the electrons is conserved
or only partially released to the test mass, {\it i.e.} the electrons
can move ballistically through the test mass. A partial conservation
of the momentum of the electrons is also obtained with the diminishing
of the number of elastic processes in a small travelled length. In
this Letter we discuss a new definition of the momentum and energy
transferred by the tunneling electrons to the test mass which is more
adequate for dealing with such situations. Some consequences relevant
for the proposed devices are finally stressed. 

The vacuum gap between the test mass and a tip put close to its
surface is schematized by a potential $V(x)$ taken to be a
one-dimensional barrier extending between points $a$ and $b$ 
(the tip is located at $x < a$ and the test mass at $x > b$). 
The force $\partial V/\partial x$ imparted by the tunneling electrons
to the two sides of the barrier mat be decomposed in two
contributions,
\begin{equation}
\frac{\partial V}{\partial x}=\frac{\partial V_1}{\partial x}+
\frac{\partial V_2}{\partial x},
\end{equation}
representing the forces inparted to the tip and to the test mass
respectively. The decomposition procedure is explained in
ref. \cite{PLA4} for two relevant shapes of $V(x)$, namely a piecewise
constant or a linearly varying potential for $a < x < b$. 
For instance, in the case of a square well barrier of 
height $V_0$ the two forces are the $\delta$-distributions 
$\partial V_1/\partial x=V_0 \delta(x-a)$ and  
$\partial V_1/\partial x=-V_0 \delta(x-b)$.
The force $\partial V_2/\partial x$ is relevant in calculating
\cite{PLA4,PLA5} the momentum current $J_p^t$ transferred to the test
mass by an electron in a tunneling eigenstate $\psi$,
\begin{equation}
J_p^t=J_p(b^+)+\int_{a^-}^{b^+} 
\frac{\partial V_2}{\partial x} \psi^* \psi dx
\end{equation}
where $J_p(b^+)$ is the momentum current $J_p(x)$,
\begin{equation}
J_p(x)=\frac{\hbar^2}{4m}\left(2 
\frac{\partial \psi^*}{\partial x} \frac{\partial \psi}{\partial x}-
\psi^* \frac{\partial^2 \psi}{\partial x^2} -
       \frac{\partial^2 \psi^*}{\partial x^2}\psi \right),
\end{equation}
evaluated inside the test mass ($x=b^+$). Eq. 2 is obtained, under
stationary conditions, from the continuity equation for the momentum
flux which translates, in a quantum mechanical framework, the second
law of dynamics
\begin{equation}
\frac{\partial \rho_p}{\partial t}+\frac{\partial J_p}{\partial x}=
-\frac{\partial V}{\partial x} \psi^* \psi
\end{equation}
and by considering only the contribution to the force due to the test
mass. Analogously, in the case of a first quantization approach use is
made of the transferred momentum squared current
\begin{equation}
J_{p^2}^t=J_{p^2}(b^+)-i \hbar \int_{a^-}^{b^+} 
\frac{\partial V_2}{\partial x} 
\left(
\psi^*\frac{\partial \psi}{\partial x}-
\psi\frac{\partial \psi^*}{\partial x}-
\right) dx
\end{equation}
where $J_p^2(b^+)$ is the momentum squared current $J_p^2(x)$,
\begin{equation}
J_p^2=i\frac{\hbar^3}{4m}
\left(
\psi^*\frac{\partial^3 \psi}{\partial x^3} - 
\frac{\partial \psi^*}{\partial x}\frac{\partial^2 \psi}{\partial x^2} +
       \frac{\partial^2 \psi^*}{\partial x^2}\frac{\partial \psi}{\partial x}-
\frac{\partial^3 \psi^*}{\partial x^3}\psi 
\right),
\end{equation}
evaluated inside the test mass. As in the case of the momentum current
eq. (5) is obtained, under stationary conditions, from the continuity
equation for the momentum squared flux
\begin{equation}
\frac{\partial \rho_p^2}{\partial t}+\frac{\partial J_p^2}{\partial x}=
i \hbar \frac{\partial V}{\partial x} 
\left(\psi^* \frac{\partial \psi}{\partial x}-\frac{\partial \psi^*}{\partial x}\psi\right)
\end{equation}

In eq. (2) and (5) the momentum and momentum squared currents
transferred to the test mass consist of two terms. The first one
represents the momentum and momentum squared current, proportional to
the energy, of the electrons moving inside the test mass. The second
one is the contribution due to the quantum mechanical scattering at
the interface between the vacuum and the test mass. Let us consider a
model of interaction in which the transferred momentum and momentum
squared currents are written, respectively as
\begin{equation}
\tilde{J}_p^t=\int_{a^-}^{b^+} \frac{\partial V_2}{\partial x} \psi^*
\psi dx,
\end{equation}

\begin{figure}[t]
\includegraphics[width=0.95 \columnwidth]{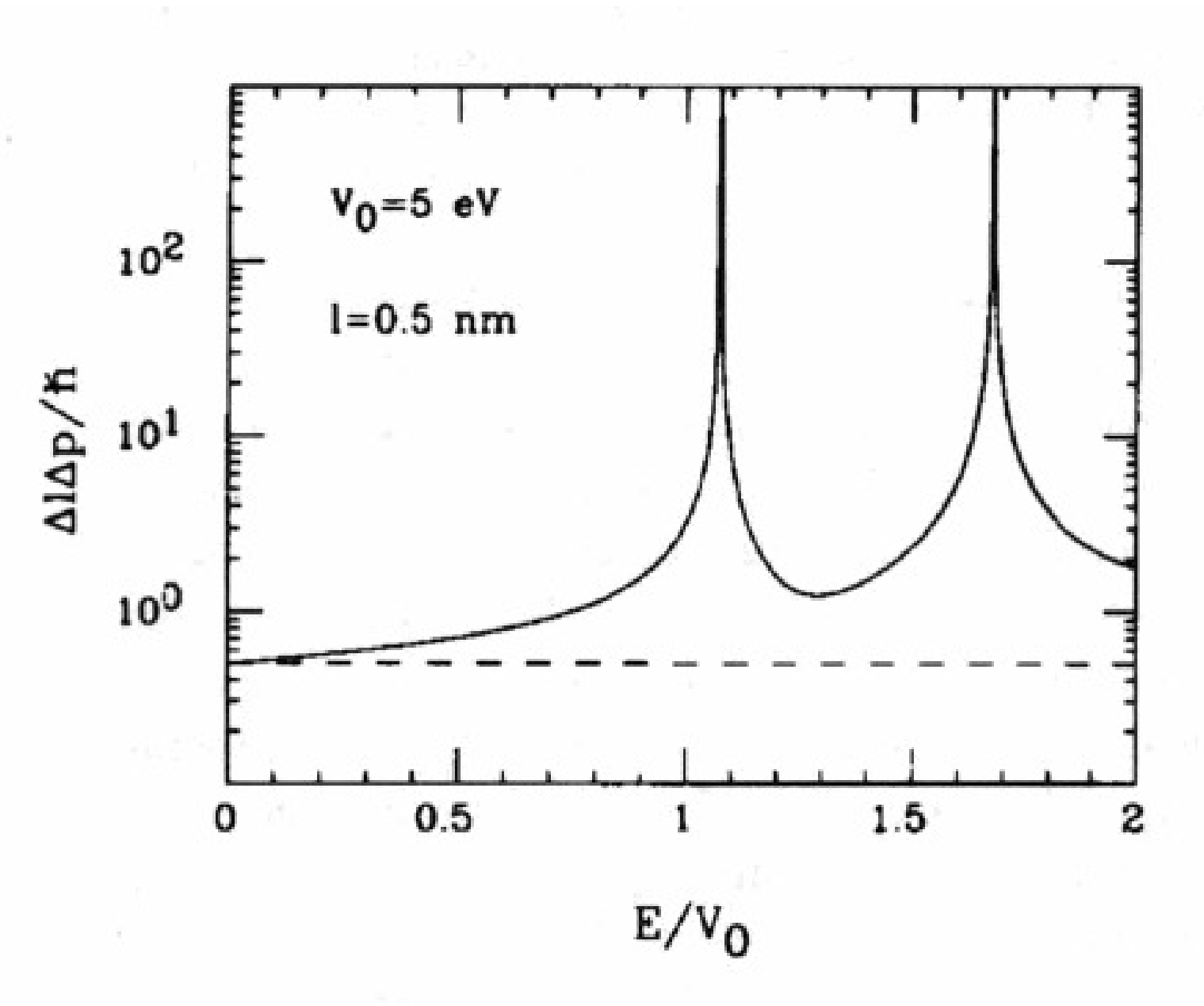}
\caption{Heisenberg uncertainty product of the test mass versus the
  energy of the incident electrons (normalized to the barrier
  potential) for the elastic (solid) and inelastic (dashed) models in
  a symmetrical rectangular barrier of height $V_0$ and width $l$.} 
\label{fig1}
\end{figure}

\begin{equation}
\tilde{J}_{p^2}^t=-i \hbar \int_{a^-}^{b^+} \frac{\partial
  V_2}{\partial x} 
\left(\psi^*\frac{\partial \psi}{\partial x}-\frac{\partial
  \psi^*}{\partial x} \psi\right) dx.
\end{equation}
The meaning of the new definition is quite clear: the exchange of
energy and momentum is only related to the presence of the interface. 
This definition is appropriate to describe a ballistic propagation of
the electrons inside the test mass (eq. (9)) with conservation of the
longitudinal momentum (eq. (8)) along the direction from the tip to
the test mass. In this situation we obtain, for $\tilde{J}_p^t$ and 
$\tilde{J}_{p^2}^t$ in the case of a rectangular barrier of height $V_0$
and width $l$ (the eigenstates $\psi$ are normalized with respect to
the wavevector), 
\begin{equation}
\tilde{J}_p^t=-\frac{1}{2 \pi} \frac{\hbar^2}{2m} T (k^2+k_0^2),
\end{equation}
\begin{equation}
\tilde{J}_{p^2}^t=-\frac{1}{2 \pi} \frac{\hbar^3}{m} T (k^2+k_0^2)k,
\end{equation}
where $\hbar k=\sqrt{2mE}$ and $\hbar k_0=\sqrt{2m(V_0-E)}$, $E$ being
the electron energy and $T=T(E)$ the transmission coefficient of the
barrier. They must be compared with the analogous expressions obtained
according to the definitions (2), (5),
\begin{equation}
J_p^t=\frac{1}{2 \pi} \frac{\hbar^2}{2m} T (k^2-k_0^2),
\end{equation}
\begin{equation}
J_{p^2}^t=-\frac{1}{2 \pi} \frac{\hbar^3}{m} T k_0^2 k,
\end{equation}

\begin{figure}[t]
\includegraphics[width=0.95 \columnwidth]{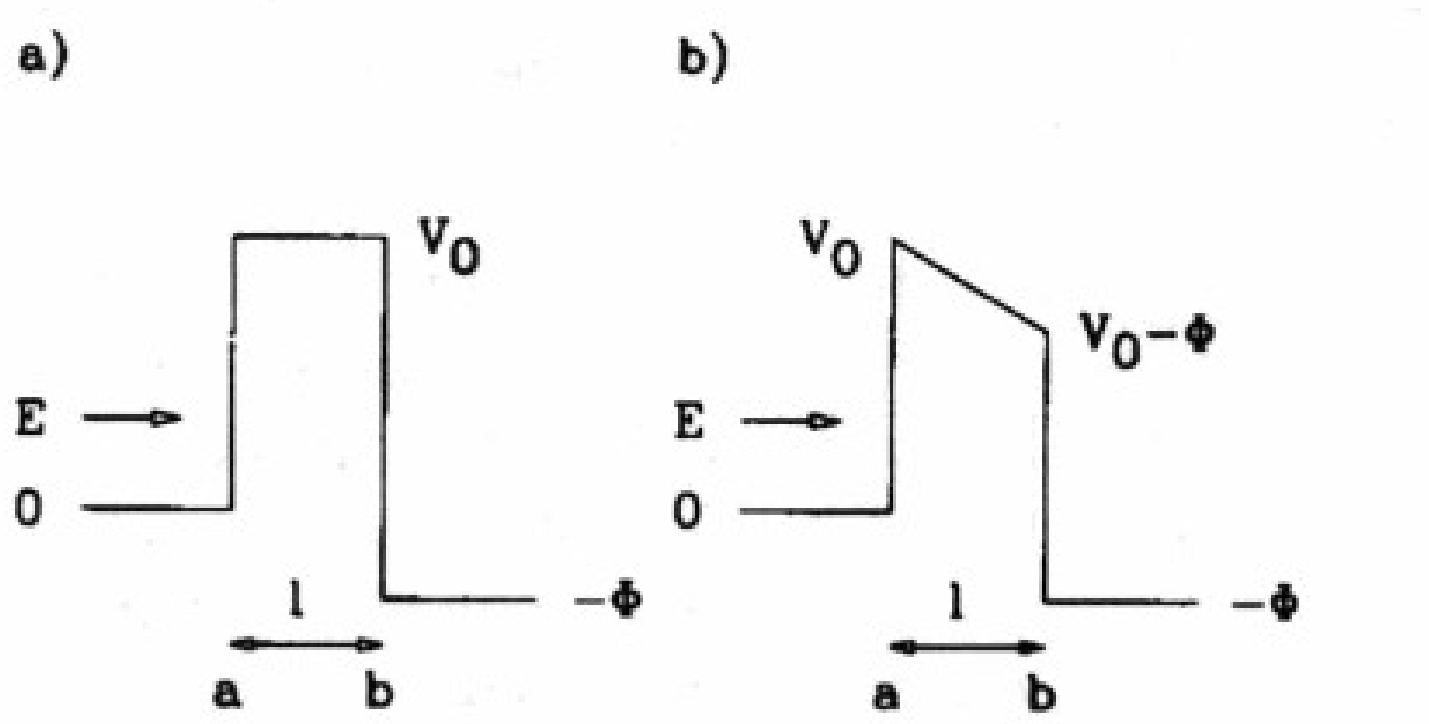}
\caption{Asymmetric rectangular barrier (a) and linearly slowing
  barrier (b). $\varphi$ represents the voltage drop between the top
  and the test mass} 
\label{fig2}
\end{figure}

We observe that, in the limit $V_0 \to 0$, $k_0 \to ik$ and 
$\tilde{J}_p^t, \tilde{J}_{p^2}^t \to 0$. On the other hand (12) and
(13) do not have the same limit, this fact expressing an exchange of
energy and momentum also in the absence of the barrier, namely a
release of these quantities to the second electrode. For this reason
we will call, in the following considerations, the first model
corresponding to (8) and (9) the {\sl elastic model}, the latter model
corresponding to (2) and (5) the {\sl inelastic model}. 

\begin{figure}[t]
\includegraphics[width=0.95 \columnwidth]{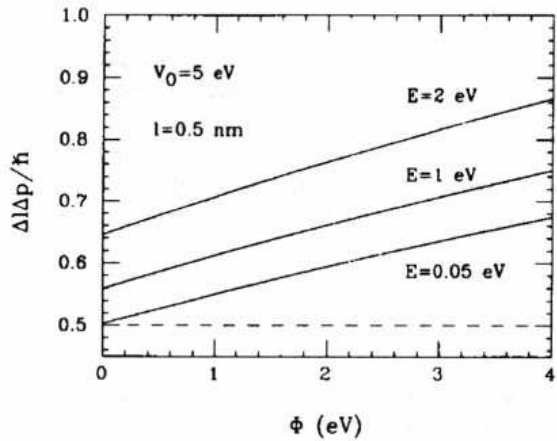}
\caption{Heisenberg uncertainty product of the test mass versus
  voltage drop for the elastic (solid) and inelastic (dashed) models
  for three different electron energies in the case of an asymmetric
  rectangular barrier. In the elastic case the uncertainty product has
  always the same value.}
\label{fig3}
\end{figure}

The evaluation of the momentum uncertainty of the test mass due to $N$
incident electrons is obtained from the mementum and momentum squared
currents \cite{PLA5} and in the elastic and inelastic models,
respectively gives
\begin{equation}
\Delta \tilde{p}^2= N \frac{\hbar^2}{4 k^2} T [4 k^2(k^2+k_0^2)+T(k^2+k_0^2)^2]
\end{equation}
and
\begin{equation}
\Delta p^2= N \frac{\hbar^2}{4 k^2} T [4 k^2 k_0^2+T(k^2-k_0^2)^2].
\end{equation}
Note that $\Delta \tilde{p}^2 \geq \Delta p^2$. The test mass shows
also a position uncertainty $\Delta l$. This arises from the
uncertainty $\Delta N$ in the number of tunneling electrons through
the dependence of the transmission coefficient on the width $l$ of the
vacuum gap \cite{PLA5},
\begin{equation}
\Delta N=\sqrt{NT(1-T)}=N |{\frac{\partial T}{\partial l}}| \Delta l,
\end{equation}
and it gives us finally the uncertainty products $\Delta l \Delta p$
for both the elastic and inelastic models. These are shown in fig. 1
as a function of the energy of the incident electrons in the case of a
rectangular barrier having $V_0=5$ eV and $l=$0.5 nm. The graph also
includes the tunneling for an electron energy greater than the barrier
height, obtained by simply replacing $k_0 \to i ik_0$ in eqs. (14) and
(15). The peaks in the curve of the elastic case are due to the
divergence in the position uncertainty in the proximity of the zeros
of the derivative of the transmission coefficient with respect to the
displacement. By taking into account the second order expansion
\begin{equation}
\Delta N=N \frac{\partial T}{\partial l} \Delta l+\frac{1}{2} N 
\frac{\partial^2 T}{\partial l^2} \Delta l^2+O(\Delta l^3),
\end{equation}
the divergence disappears but in this case the transduction of the
displacement is not linear. So one should avoid such conditions for a
proper working of the device. We observe that these points are always
in the regime of energy higher than the height of the
barrier. Moreover, the divergences disappear considering, instead of
an energy eigenstate for the tunneling electron, a more realistic wave
packet. 

\begin{figure}[t]
\includegraphics[width=0.95 \columnwidth]{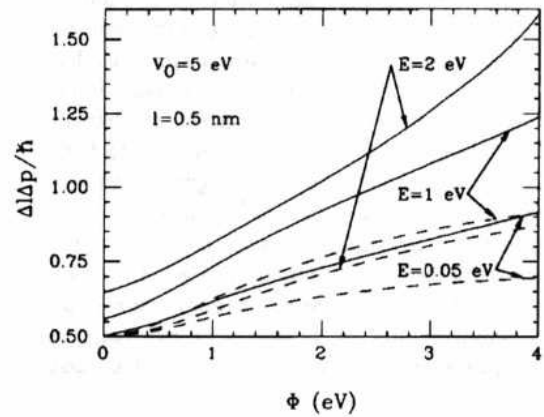}
\caption{Heisenberg uncertainty product of the test mass versus
  voltage drop for the elastic (solid) and inelastic (dashed) models
  for three different energies in the case of a linear slowing barrier.}
\label{fig4}
\end{figure}

We have repeated the calculations of the uncertainty product for an
asymmetrical rectangular barrier (fig. 2a) and a linearly slowing
barrier (fig. 2b). Some results are shown, respectively, in figs. 3 
and 4 as a function of the drop voltage across the barrier for
different values of the incident energies. In both the cases the
elastic model predicts higher sensitivity to the drop voltage with
respect to the inelastic model. 

\begin{figure}[b]
\includegraphics[width=0.60 \columnwidth]{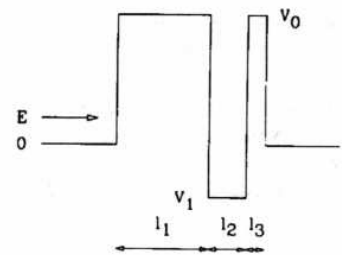}
\caption{Double barrier potential for the tunneling through an
  adsorbed atom.}
\label{fig5}
\end{figure}

The situation corresponding to electron tunneling through a double
barrier potential (fig. 5), was already studied in ref. \cite{PLA4} as
a schematization of an atomic impurity near the surface of an
electrode. In fig. 6 the momentum flux transferred to the test mass is
shown versus the electron energy for both the elastic and inelastic
models. The behaviour of the two curves is very similar and in both
the cases the momentum flux goes from negative to positive values for
increasing energy of the tunneling electrons crossing the zero for an
energy roughly corresponding to the peak of maximum transmission. When
the energy of the electron is small, the two curves are almost
coincident. 

\begin{figure}[t]
\includegraphics[width=0.95 \columnwidth]{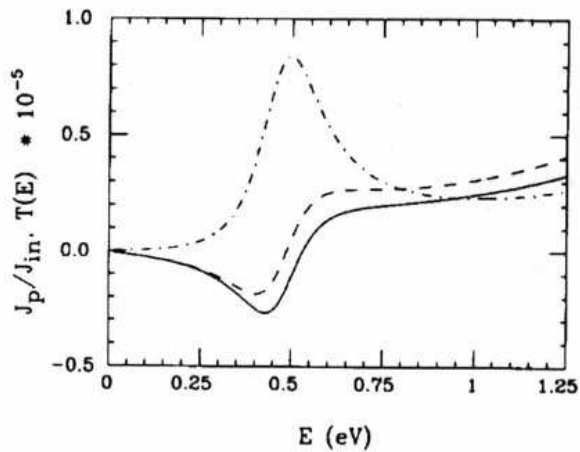}
\caption{Transmitted momentum current normalized to the incident
  electron current for the elastic (solid) and inelastic (dashed)
  models and transmission coefficient (dot-dashed) versus the energy
  of the electrons tunneling through the double barrier of fig. 5 with 
$V_0=$ 4 eV, $V_1=$-2.1 eV, $l_1=8 \AA$, $l_2=2 \AA$ and $l_3=1.2 \AA$.}
\label{fig6}
\end{figure}

To understand what the physical situations are in which the elastic
model is more adequate to describe the electron barrier interaction we
recall that both the proposed applications in refs. \cite{PLA4,PLA5},
because of the need for a momentum detection, are meaningful only if
other sources of mechanical noise, like Brownian motion, are made
negligible. This is obtained if the devices operate at very low
temperature, of the order of 1-10 mK. In this case the electron-phonon
coupling, proportional to the temperature, is negligible with respect
to the electron-electron scattering. This last contribution has
already been investigated in bulk metals and both models and
measurements are in agreement with an increase of the mean free path
of the electrons $\lambda(E)$ when their energy is below 20-30 eV,
this last value depending upon the specific material. At low energy
the behaviour of $\lambda(E)$ follows approximately the law
\cite{PLA6}

\begin{equation}
\lambda(E)=A/E^2+B/\sqrt{E}
\end{equation}
where $A$ and $B$ are empirically known. In the range which is of
interest for tunneling of electrons, {\it i.e.} $10^{-1}$-1 eV, mean
free paths of the order of $10^4-10^5 \AA$ are estimable. Another
possibility is to consider the test mass to be a semiconductor
crystal. In this case an electron mobility of 10 m${}^2$/V s can be
achieved \cite{PLA7} which again gives a mean free path of $10^5 \AA$
for electrons of energy equal to 0.1 eV. In both the cases, despite
the crude approximations, we have a mean free path of the same order,
or more, of the size of the micromachined test masses to be used in
the devices. In a very low energy regime we have shown that the
results of the two models discussed here are almost
identical. However, a range of energies in which tunneling happens and
in which the two models give different predictions exists, according
to the graphs in figs. 1, 3, 4 and 6. Therefore, we conclude that the
elastic model has to be taken into account as a more adequate tool for
the design of small-size, micromechanical devices based upon detection
of momentum exchange in a tunneling probe.

\acknowledgments
We thank F. Sacchetti for useful discussions and B.S. Waller for a
critical reading of the manuscript.

\end{document}